\documentclass[preprint,letterpaper,showpacs,preprintnumbers,amsmath,amssymb]{revtex4}
\usepackage{dcolumn}
\usepackage{graphicx}
\usepackage{rotating}
\input{epsf.tex}

\begin{document}

\title{Nitrogen clusters inside $C_{60}$ cage and new nanoscale energetic materials }

\author{Hitesh Sharma$^{1,2}$}%
\author{Isha Garg $^{1}$}%
\author{Keya Dharamvir$^{1}$}
\author{V. K. Jindal$^{1}$}\thanks{Corresponding author: Fax: +91 1722783336.
E-mail address: jindal@pu.ac.in (V. K. Jindal)}
\affiliation{%
$^{1}$Department of Physics, Center of Advanced Studies in Physics,
\\Panjab University, Chandigarh, India-160014\\
$^{2}$Department of Applied Sciences, RBIEBT, Sahauran,
Mohali,Punjab-140104}

\date{\today}

%\input{epsf.tex}

\begin{abstract}
We explore the possibility to trap polynitrogen clusters inside
$C_{60}$ fullerene cage, opening a new direction of developing
nitrogen-rich high energy materials. We found that a maximum of 13
nitrogen atoms can be encapsulated in a $C_{60}$ cage. The nitrogen
clusters in confinement exhibit unique stable structures in
polymeric form which possess a large component of ($\sim$ 70-80$\%$)
single bond character. The $N_{n}$@C$_{60}$ molecules retain their
structure at 300K for n$\leq$12. The Mulliken charge analysis shows
very small charge transfer in $N@C_{60}$, consistent with the
quartet spin state of N. However, for 2$<$n$<$10, charge transfer
take place from cage surface to $N_{n}$ compounds and inverse
polarization thereafter. These nitrogen clusters when allowed to
relax to $N_{2}$ molecules which are triply bonded are capable of
releasing a large amount of energy.
\end{abstract}

\pacs{72.80.Rj, 31.15.ae, 33.15.-e, 85.65.+h}

\maketitle
\section{INTRODUCTION}
In recent years, interest in polynitrogen or polymeric nitrogen has
drawn considerable attention not only because of its theoretical
significance but also its possible application as high energy
density material (HEDM)\cite{r1,r2}. Polynitrogen molecules are
formed by a combination of lower order bonds and while decaying into
dinitrogen ($N_{2}$) molecules, they release enormous amount of
energy and are environmentally safe.
\par
Over last few decades, there have been consistent efforts to predict
new exotic forms of all-nitrogen molecules using various
experimental and theoretical techniques\cite{r3,r4,r5,r6}. The
higher members of nitrogen cluster family ($N_{n}$ for n$>$3) are
unstable in the free space and therefore present a challenge to
their synthesis. In 1999, Christe \emph{et al.}\cite{r8} synthesized
successfully a $N_{5}^{+}$ salt $N^{{+}}AsF^{-}$ which is third
readily accessible homonuclear polynitrogen species. Subsequently,
the same group reported the stability of $N_{5}^{+}SbF_{6}^-$ up to
$70^oC$ and its relative insensitivity to impact \cite{r8a}. This
discovery opened a new dimension to explore neutral polynitrogen
compounds and basis for the first synthesis of stable nitrogen
allotropes.
\par
Nitrogen rich compounds such as nitrogen hydrides $(N_{n}H_{m})$
(n$>$m) could also be used as starting materials to produce $N_{n}$
species, since $N_{4}H^{+}$, $N_{4}H_{6}$ and $N_{4}H_{4}$
\cite{r8b,r8c,r8d} species are stable and their subsequent
deprotonation may lead to the formation of neutral $N_{4}$.
Similarly, the oxidation of nitrogen oxide $N_{4}O$ which is stable
and has been observed experimentally \cite{r8e}, would also lead to
the formation of $N_{4}$. Further, it has been suggested that higher
$N_{n}$ could be synthesized via direct or indirect excitation of
electronic states in liquid and solid nitrogen followed by collision
addition of a ground state N and excited $N_{n-1}$ or $N_{n-2}$
thereby forming higher $N_{n}$ compounds. For example, by combining
different $N_{n}$ moieties such as $N_{5}$ and $N_{3}$ might lead to
formation of $N_{8}$ species \cite{r8f} and $N_{5}^{-}N_{5}^{+}$ may
form $N_{10}$ \cite{r8g, r8h}. In 2002, Cacace \emph{et al.}
\cite{r9} demonstrated the existence of the tetranitrogen molecule
($N_{4}$) as a metastable species whose life time exceeds 1 $\mu$s
at 298K. The identification of $N_{4}$ represents the first addition
in nearly half a century to the family of the polynitrogen
molecules. Hammerl and Klapotke \cite{r10} also reported a combined
theoretical and experimental NMR study on nitrogen rich compounds.
At the most fundamental level, a molecule will exist only if it has
high enough energy barrier that keeps it away from dissociation.
\par
This possibility has been explored theoretically. Various novel
polynitrogen species have been studied in variety of forms such as
cyclic, acyclic or caged conformations. The species investigated
include ionic clusters \cite{r10a,r10b,r10c}, cylinders \cite{r10d},
cages \cite{r10e,r10f,r10g,r10h}, nanoneedles \cite{r10i}and
helices\cite{r10j}. Isomers of smaller systems $N_{7}$ \cite{r10k},
$N_{10}$ \cite{r10l} and $N_{12}$ \cite{r10m} have also been
reported theoretically in free space. Majority of these studies have
focused on the evaluation of these novel forms of molecular
nitrogen. However, Wang \emph{et al.}\cite{r10n} has described the
possible synthesis route for all-nitrogen systems. Dixon \emph{et
al.}\cite{r10o} has successfully demonstrated the ability of the
theory to predict the stability of such compounds. In spite of these
efforts, so far no higher members (n$>$4) of polynitrogen family in
free space have been synthesized experimentally . The present
difficulties in synthesizing the polynitrogen molecules motivates us
to look for new avenues to realize the success of polynitrogen
molecules as a possible HEDM.
\par
Further, the new polymeric forms of nitrogen have also been explored
in extreme conditions of high pressure and temperature. The
application of high pressure to nitrogen system leads to destruction
of covalent bonds and may lead to formation of intermediate
polymeric network of single and double bonded atoms which has been
verified experimentally \cite{r10p,r10q,r10r,r10s}. Using \emph{ab
initio} calculations, other forms of polymeric nitrogen such as
Black Phosphorous (BP) \cite{r10t}, A7 \cite{r10u} ,
metallic\cite{r10v}, simple tetragonal phase \cite{r10v} and Cubic
Gauche (CG) form \cite{r10s} have been proposed at extreme
conditions. The CG form has also been observed experimentally
\cite{r11}. Therefore, the success of \emph{ab initio} studies
\cite{r11a} in predicting new polymeric phases of nitrogen in
extreme conditions motivates us to study polynitrogen in other
physical environment such as in confinement.
\par
More recently in 2008, nitrogen has been predicted to exist in
polymeric form ($N_{8}$) in confinement of nanostructures i.e.
Carbon Nanotubes. They proposed $N_{8}$ stabilizes as a chain which
is stable up to room temperature \cite{r12}. Therefore, apart from
extreme condition of high temperature and pressure, confinement
offers an alternate environment for stabilization of polynitrogen
compounds. Since the environment of $C_{60}$ has been extensively
studied by our group \cite{r12a,r12b,r12c,r12d} as well as others
\cite{r12e,r12f} it offers a new possibility for polynitrogen
encapsulation. The $C_{60}$ cage has already been explored for
endohedral doping with alkali metals \cite{r13}, transitional metals
\cite{r14}, non-metals \cite{r15} and noble gases \cite{r16}. Among
all endohedral complexes, nitrogen-doped $C_{60}$ are quite
interesting and needs further investigation. As per our knowledge,
$N@C_{60}$ is the only endohedral complex which have been explored
extensively theoretically \cite{r12e,r12f,r17,r18} as well as
experimentally \cite{r19}. However, $N_{2}@C_{60}$ was successfully
synthesized using pressure heating \cite{r19a} and one isolated
study of its stabilization energy has been recently reported
\cite{r19b}. Till date, the higher members of nitrogen family
(n$>$2) have not been explored at endohedral position in $C_{60}$.
New techniques have been developed to deal with nitrogen (as
impurity) in nanostructures such as chemical vaporization deposition
(CVD) method \cite{r19d} which is capable of synthesizing nitrogen
doped carbon nanotubes. And recently in 2008, $C_{57}N_{3}$ was
reported using surface-catalyzed cyclodehydrogenation from aromatic
precursors \cite{r19e} and further expressing a strong possibility
of synthesizing other nitrogen-based hetero and endohedral
fullerenes. Keeping this in view, further efforts are needed to be
invested in both experimental and  theoretical sides to help in
designing new ways for generating $N_{n}$ compounds and understand
their stability in different environments.
\par
In this paper, we report the results of our \emph{ab initio}
calculations on $N_{n}@C_{60}$ for n$\leq$16, which describes the
novel possibility of stabilizing the higher members of all-nitrogen
molecules in confined environment of $C_{60}$. We intend to show
that $C_{60}$ can act as an ideal candidate to trap nitrogen in
polymeric form and can be used as a possible high energy density
material. The present study would provide some valuable information
for synthesizing more stable all-nitrogen clusters to develop the
novel materials for future.
\section{COMPUTATIONAL DETAILS AND VALIDATION}

\par
We have used the Spanish Initiative for Electronic Simulation with
thousands of atoms (SIESTA) computational code\cite{r20} which is
based on numerical atomic orbital density functional theory method
\cite{r21,r22,r23}. We have used $1s^{2}$, $2s^{2}$, $2p_{x}^{1}$,
$2p_{y}^{1}$, $2p_{z}^{1}$ $(4S_{3/2})$ configuration of atomic
nitrogen corresponding to spin quartet state which is consistent
with the experimental observation \cite{r19}. The calculations are
carried out using generalized gradient approximation (GGA) that
implements Becke gradient exchange correlation functional
(BLYP)\cite{r24}. Core electrons are replaced by non-local,
norm-conserving pseudopotentials factorized in the Kleinman-Bylander
form \cite{r25}, whereas valence electrons are described using
linear combination of numerical pseudo atomic orbitals of the
Sankey- Niklewski type \cite{r26} but generalized for
multiple-$\zeta$ and polarization functions. In this work, we have
used a split valence double-$\zeta$ (DZP) basis set with energy
shift equal to 100 meV. The real space cutoff grid energy is taken
as 200 Ry. Periodic boundary conditions have been considered using
simple cubic shell of 2 nm that is large enough to avoid any
significant spurious interactions with periodically repeated images.
\par
We first performed test calculations on the C$_{60}$ and $N_{2}$
molecules. In $C_{60}$, we found the C=C and C-C bond distances as
0.140 nm and 0.146 nm in agreement with experimental values
\cite{r27}. The ionization potential and electron affinity were
found to be 6.9 eV and 2.70 eV, in good agreement with experimental
values of $7.5 \pm 0.01$\cite{r28} and $2.689 \pm 0.008$ \cite{r29}
respectively. In $N_{2}$ molecule, the N-N bond distance and
ionization potential are found to be 0.112 nm and 15.43 eV which are
in good agreement with experimental values of 0.115 nm and 15.60 eV
respectively \cite{r30}.
\par
We now proceed with the investigation of nitrogen clusters inside
$C_{60}$. We started with different initial atomic configurations of
$N_{n}$ for n = 1-16. The following considerations were taken into
account while starting and determining the final structure for
particular configuration. (a) Nitrogen atoms were placed
sufficiently away from the walls of $C_{60}$ so that there is no
likelihood to form any C-N bond. (b) Nitrogen atoms were allowed to
form bonds among themselves and make polynitrogen compounds. (c)
Each structure was allowed to relax till forces on each atom
converges up to 0.001 eV/$\AA$. (d) The structure with minimum
energy is considered as the final stable polymeric form of nitrogen.
(e) All the optimized nitrogen clusters inside $C_{60}$ were also
relaxed in free space at 0K. (f) The structures of $N_{n}$@$C_{60}$
were also relaxed at room temperature (300K) to confirm their
stability (g) The spin polarized calculations were performed on
final stable polymeric $N_{n}@C_{60}$ to determine the spin states
(h) Also harmonic vibrational frequency analysis on all optimized
structures was performed to further ensure their ground state
geometries.
\section {RESULTS AND DISCUSSION}
The optimized ground state structures of polynitrogen compounds
inside $C_{60}$ are shown in Figure 1-3. We started our calculation
with single nitrogen, the nitrogen was placed at different positions
inside $C_{60}$ and the SCF total energy isosurface is explored
throughout region inside $C_{60}$.  We found the atomic nitrogen at
center as stable structure with small zero point vibrations. Since
the atomic nitrogen is highly reactive so if we place nitrogen at a
distance of more than 0.150 nm from center of cage, it has tendency
to form a weak covalent bond of the order of 0.154 nm with two
adjacent carbon atoms of hexagon-hexagon interface of $C_{60}$ inner
surface\cite{r31}. When two nitrogen atoms are placed inside
$C_{60}$ at different positions, they form $N_{2}$ molecule which is
found to stabilize at the center of the $C_{60}$. The resultant
structure is $N_{2}$ molecule with N$\equiv$N bond distance of 0.112
nm which is same as in free space.
\par
In $N_{3}@C_{60}$,  $N_{3}$ in linear form with $D_{\infty h}$
symmetry is found to be stable at the center with N-N bond length of
0.119 nm. However, the triangular structure $D_{3h}$ is found to be
another stable isomer inside $C_{60}$. Interestingly, most of
initial $N_{3}$ configurations including the combination of $N_{2}$
and single N at distance results in an isosceles triangular
structure with bondlengths of 0.133 nm and 0.158 nm and bond angles
53.5$^{o}$ and 73.1$^{o}$ respectively. The energy difference
between both the structures is of the order of 0.9 eV. $N_{4}$ is
found to exist as two dimers of $N_{2}$ with bond length of 0.111 nm
and placed at a distance of 0.226 nm symmetrically across the center
of cage. The arrangement of two $N_{2}$ molecules on either side of
center of $C_{60}$ may be explained on the basis of extra stability
of $N_{2}$ and its reluctance to form lower order bond. The
structure with $T_{d}$ and $D_{2h}$ symmetry are found to be the
other stable isomers. However, in free state $N_{4}$ has been
predicted to be stable in $C_{2h}-T$, $T_{d}$ and $D_{2h}$
symmetries respectively \cite{r32}. $N_{5}$ is found to be stable at
center in perfect pentagon shape having $D_{5h}$ symmetry with each
bond length of 0.132 nm and bond angle of 108.0 $^o$. We tried
various initial configurations of $N_{5}$ (including dimers and
single atom). It was observed that all the configurations lead to
the formation of pentagon structure which may be due to confined
environment of $C_{60}$.
\par
For n $\geq$ 6, different combinations of N, $N_{2}$, $N_{3}$ and
higher polynitrogens were considered as separate substitutional
identities in $C_{60}$. We found interesting pattern in the
arrangement of nitrogen atoms in polynitrogen compounds, which are
very stable and are placed symmetrically across the center. The
other observed isomeric structures, though stable, are found
distinctly less stable by an energy difference of $\sim$ 0.5-2.0 eV.
The minimum energy configuration structure of $N_{6}$ comes out to
be a boat-shaped cyclic ring with bond lengths of the order of
0.128-0.135 nm and bond angles 108.8$^o$ and 122.5 $^o$. In free
state $N_{6}$ is found to be stable in hexaazadewarbenzene structure
with $C_{2v}$ symmetry which is also one of the stable isomers in
confinement \cite{r33}.
\par
$N_{7}$ is found to exist as an assembly of a pentagon with bond
lengths 0.128-0.132 nm  and a dimer having bond length 0.111 nm as
two separate units. The presence of $N_{5}$ as an independent
identity is responsible for lowering the energy of $N_{7}@C_{60}$
indicating its extra stability . However, one closed structure
involving $N_{5}$ and  $N_{2}$ do exist as a close energy isomer as
shown in Figure1. $N_{8}$ is found to exist in two planar rings
having distinct bond lengths of 0.127, 0.138, 0.145 and 0.138 nm
which are connected together with a weak N-N bond of 0.157 nm.
$N_{9}$ exhibits a stable structure with $N_{5}$ and $N_{4}$ planar
rings connected together with a bond of 0.148 nm. The bond lengths
for both the rings are in the range 0.127-0.150 nm. $N_{10}$ forms a
structure with two planar symmetrical pentagon rings on either side
of center of $C_{60}$ ($D_{5h}$ symmetry) connected to each other
with a single bond of 0.150 nm. $N_{11}$ polynitrogen compound is
also stable in ring structure with one planar pentagon and one
non-planar hexagon connected by a bond distances 0.142 and 0.145 nm
respectively. $N_{12}$ stabilizes itself symmetrically into two
non-planar boat shaped hexagons connected by N-N bond of 0.139 nm.
\par
For n$>$12, the polynitrogen compounds do not show any symmetry in
distribution of nitrogen atoms due to the availability of less
space. They make complex yet stable structures. $N_{13}$ is found to
exhibit a geometry with one planar pentagon connected to a
non-planar closed $N_{8}$ ring with significant distortion on
$C_{60}$ cage. $N_{14}$ and $N_{15}$ compounds make distorted
endohedral complex with nitrogen undergoing chemisorption at the
surface breaking the surface C-C and C=C bonds and making C-N bond
of 0.136 nm. This leads to the formation of heptagon in the cage
which results in very large distortion and the maximum and minimum
diameter is observed to be 0.760 nm and 0.730 nm respectively. At
n$=$16, the chemisorption of three nitrogen atoms results in the
opening of the $C_{60}$ cage.
\par
The encapsulation energy is defined as
\begin{equation}
\Delta E = E[N_{n}@C_{60}]-E[C_{60}]-E[N_{n}] ,
\end{equation}
where $E[N_{n}@C_{60}]$ is the energy of the complex, $E[C_{60}]$ is
the energy of an isolated $C_{60}$ and $E[N_{n}]$ is the energy of
an isolated $N_{n}$ cluster. These energies have been calculated for
all the structures and are tabulated in Table I. We would like to
add here that the energy of $N_{n}$ cluster in isolation is
different from the $N_{n}$ in confinement as their structures are
very different. These constrained polynitrogen molecules
disintegrate into smaller clusters when relaxed in isolation. The
encapsulation energy of $N_{2}@C_{60}$ is found to be 13.8 kcal/mol,
which is in agreement with the value of 9.11 kcal/mol reported using
B3LYP \cite{r19b}. However we would like to mention here that B3LYP
does not describe the non bonding systems accurately \cite{r19c}.
\par
The encapsulated polynitrogen clusters results in distortion of
icosahedral symmetry of the $C_{60}$ thereby altering the electronic
properties of $N_{n}@C_{60}$ significantly. The HOMO-LUMO gap
variation as a function of number of nitrogen atoms shows gradual
decrease with higher values for n= 2, 4, 6, 8, 10. To visualize this
we have plotted the Kohn Sham energy levels for $N_{n}@C_{60}$
complexes using DZP basis in Figure 4. Also electronic density of
states (DOS) for the structures are shown in Figure 5-6. We carried
out spin polarized calculations on all the optimized geometries.
Multiplicities defined as (2s+1), where s is the spin of the system,
are calculated from the difference of spin up and spin down
electrons (Q up down) and are tabulated in Table II.
\par
For $N@C_{60}$ we found a spin quartet state when the N atom is
placed at the center of the $C_{60}$ cage i.e N atom retains its
atomic nature, which is consistent with the experimental ESR or EPR
spectra of $N@C_{60}$ \cite{r12e,r17,r17a}. However when the N atom
moves toward the surface and makes a covalent bond with the cage, it
shows a spin doublet state consistent with experimental results
\cite{r31}. All the structures with even number of nitrogen atoms
show spin doublet states (s=1/2) whereas when odd number of nitrogen
atoms are present they show spin singlet (s=0) state except for n=13
which shows a spin quartet state (s=3/2).
\par
The net charge on the polynitrogen compounds and the fullerene cage
atoms are investigated by Mulliken charge analysis. We would like to
mention here that charge transfers in SIESTA are dependent on the
choice of basis set. Therefore, we carried out test calculations
using  SZ (single zeta), DZ (double zeta) and DZP (double zeta with
polarization) (with energy shift ranging from 50 meV -250 meV) basis
sets on $N@C_{60}$. It was observed that using SZ basis N shows a
small charge loss of 0.04 electrons to $C_{60}$ cage, whereas using
DZ and DZP basis N shows a small charge gain from the cage. The
magnitude of charge transfer is found to be higher when DZ is used
compared to that using DZP. Within DZP basis set as we increase the
energy shift the mulliken charge on N tends to increase marginally
from 0.027 (energy shift 50 meV) to 0.040(250 meV). Our work on
azafullerenes using DZP basis set produced reliable charge transfers
in $C_{60-n}N_{n}$ \cite{r34} and it reproduced the experimental
parameters of $C_{60}$ and $N_{2}$ very accurately.
\par
For $N@C_{60}$, the small charge transfer of 0.04 electrons from the
cage to N at center is found to be consistent with spin quartet
state as obtained using spin polarized calculations. However, there
are conflicting claims regarding charge transfers in $N@C_{60}$
using various \emph{ab initio} calculations. Kobayashi \emph{et al.}
\cite{r35} using 6-31G(d) basis set and B3LYP functional has
reported slightly large spin density at N in $N@C_{60}$ than in
$N@C_{70}$ consistent with observed hyperfine coupling constants. Lu
\emph{et al.} \cite{r36} using user defined basis set (considering
2s and 2p for C and N) and carrying discrete variational local
density functional calculations had reported charge transfer of 0.11
from $C_{60}$ cage to N atom at the center. Further, Mauser \emph{et
al.} \cite{r31} using semi empirical and DFT(B3LYP) and Greer
\cite{r37} using TZVP spin-unrestricted DFT optimized geometry
within DZP basis sets have predicted no charge transfers between N
and $C_{60}$. For n=2-10, we found charge transfer from cage to
$N_{n}$, which may be explained on the basis of higher values of
electronegativity and ionization potential of N than C atoms. The
Mulliken charge calculation suggests electron charge transfer
ranging between 0.05 to 0.31 from carbon atoms to the polynitrogen
compounds inside up to n=10 except for $N_{2}$ which shows inverse
nature. For n$>$10, charge transfer of 0.05-0.47 electrons takes
place from nitrogen to carbon atoms of fullerene. This reversal can
be attributed to availability of less space and overlapping of
orbitals which may be further responsible for destabilizing the
polynitrogen compounds beyond n=13. As per our information, there
are no references for comparison to any theoretical work on
polynitrogen clusters in $C_{60}$ till date.
\par
From the structural analysis of all the $N_{n}@C_{60}$ structures
and their isomers it may be concluded that N atoms have preferential
sites for its stabilization inside fullerenes. Moreover, due to
confinement the nitrogen atoms are forced to form lower order bonds.
We found that the threshold size of polynitrogen compounds to exist
inside as stable molecule is $N_{13}$. Any further addition of
nitrogen inside leads to the formation of C-N bond and heptagon
involving nitrogen atom on the surface. All the nitrogen clusters in
$N_{n}@C_{60}$ for n$>$4 has 70-80 $\%$ single bond character which
confirms the polymeric nature of nitrogen compounds as shown in
Figure 7. The structural distortion expressed in terms of elongation
defined as $ \epsilon=(\frac{R}{R_{0}}-1)$ where R and ${R_{0}}$ are
the average radii of expanded and pure $C_{60}$ is shown in Figure
8. The percentage elongation shows a linear increase with the number
of nitrogen atoms up to $N_{15}$ except for $N_{2}$ which shows
small contraction in $C_{60}$ cage. The maximum threshold elongation
is found to be $\sim$ 6$\%$ for $N_{15}$. Indeed, this elongation
leads to estimation of extra pressure. Successive addition of
nitrogen atoms exert this extra pressure from within when their
number is increased beyond 2, till the stability limit of $C_{60}$
is reached.
\par
Further, the stability of each structure is verified by displacing
the nitrogen atoms up to 0.5 nm and observing the result. We find
that after relaxing the structure , it regains its optimized form.
Moreover, harmonic vibrational frequency analysis is carried out on
all the ground state structures and the absence of any imaginary
frequency value further verifies the true energy minima. The ground
state geometries for polynitrogen $N_{n}$ compounds, when relaxed in
free space, leads to fragmentation with large number of $N_{2}$
molecule units. This suggests that the polynitrogen compounds are
forced to remain stable inside due to confinement and any mechanism
triggering the breaking of $C_{60}$ cage may lead to the release of
enormous amount of energy. The energy released can be approximately
estimated based on the energy of $N_{n}C_{60}$ and $C_{60}$ and
equivalent ($\frac{1}{2}$n of $N_{2}$ molecules). Assuming all n
atoms of N in $N_{n}C_{60}$ clusters are singly bonded (with energy
158.9 kJ/mole), the energy released on formation of $N_{2}$
molecules comes out to be around 3-6 kcal/gm of nitrogen atoms. This
is several times larger than the energy release in the case of TNT.
\par
We performed constant-temperature constant-volume molecular dynamics
for $N_{n}@C_{60}$ at a temperature of 300K with time period 10 ps
with interval of 1 fs. It was observed that for $n\leq$12 the
polynitrogen structures remain intact and do not show any
significant deviation from their structures at 0K which demonstrates
the stability of polymeric nitrogen structures. But for $n\geq$13,
the $C_{60}$ cage opens up due to formation of C-N bonds as a result
of thermal vibrations. Interestingly, our preliminary calculations
on addition of few water molecules in $N_{n}@C_{60}$ for n=8-10,
indicates that the polynitrogen dissociates resulting in breaking of
$C_{60}$ cage.
\par
The experimental synthesis of these novel hybrid systems would be a
challenging task. Owing to the instability of $N_{n}$ clusters in
free space, the most feasible and promising way is to trap them in
some nanostructures such as fullerenes, nanotubes and transition
metal complexes. It appears to us that $N_{n}@C_{60}$ could be
synthesized by nitrogen enrichment using prolonged ion implantation
of N ions or pressure heating \cite{r19a}. It is well known that
laser ablation of graphite can produce copious amounts of $C_{60}$.
When graphite, intercalated with desired specie of atom/ molecule is
laser ablated, endohedral $C_{60}$ results. Therefore, even if
short-lived $N_n$ clusters could be intercalated before laser
ablation, there is good possibility of their getting trapped inside
$C_{60}$ cage.
\section{Conclusions}
The polynitrogen compounds formed inside $C_{60}$ are found to be
quite stable in confinement. The energy difference of the optimized
ground state geometries to the isomeric states is found to be
significant, thereby indicating clear choice of ground state
structures. The spin states of $N_{n}@C_{60}$ are found to be
quartet for \textbf{n}=1 and \textbf{n}=13, doublet for other odd
\textbf{n} and singlet for even \textbf{n}. The reversal of
polarization and the space available inside $C_{60}$ may be
responsible for determining the amount of nitrogen encapsulation
inside fullerene. The maximum number of nitrogen atoms that can be
encapsulated inside $C_{60}$ is 13, which can form stable structure
at 0K. The endohedral molecules $N_{n}$@$C_{60}$ for $n\leq12$, are
found to retain their structure at room temperature. This suggests
the strong possibility of synthesizing polynitrogen compounds in
confinement which could be triggered with suitable combinations of
oxygen and hydrogen for extracting tremendous amount of energy. The
amount of energy density from conversion of lower order bonds to
triple bonds comes out to be about three times in contrast to the
conventional chemical energetic materials.
\\\textbf{{ACKNOWLEDGEMENTS}}
\\Authors are thankful to SIESTA group for providing their
computational code and greatly acknowledge the computational support
provided by Prof. D. G. Kanhere, University of Pune, Pune. Isha Garg
is thankful to University Grants Commission, New Delhi
for providing financial support. \\
\\

\begin{table}
\textbf{Table I: The encapsulation energy in eV defined as delta E =
$E _{complex}$-$E_{C_{60}}$-$E_{N_{n}}$, where $E_{N_{n}}$ is the
energy of the $N_{n}$ compounds in isolation in free space. Note the
energy considered here are of theoretically lowest energy structures
which are very different from structures in confinement and do not
exit in free space for n$>$3.  }\\

\begin{tabular}
{|c|c|c|c|c|c|c|}\hline
  $N_{n}@C_{60}$ & Total Energy (eV)& Encapsulation energy (eV)  \\
  \hline
  $N@C_{60}$& -9490.2 & -2.4  \\
  &         &         \\
  $N_{2}@C_{60}$  & -9766.0 & -0.6 \\
 &         &               \\
  $N_{3}@C_{60}$ & -10030.5 & 0.2\\
 &         &                \\
  $N_{4}@C_{60}$ & -10301.0& -3.5 \\
 &         &                \\
  $N_{5}@C_{60}$  & -10566.7 & 2.9  \\
 &         &               \\
  $N_{6}@C_{60}$ & -10831.2 & 11.8  \\
 &         &               \\
  $N_{7}@C_{60}$  & -11097.8 & 10.7\\
 &         &               \\
  $N_{8}@C_{60}$  & -11360.1 & 9.8 \\
 &         &              \\
  $N_{9}@C_{60}$ & -11624.6 & 16.4\\
 &         &               \\
  $N_{10}@C_{60}$  & -11889.4 & 25.5 \\
 &         &              \\
  $N_{11}@C_{60}$  & -12152.6 & 33.9  \\
 &         &                \\
  $N_{12}@C_{60}$ & -12414.9 & 37.7 \\
 &         &                \\
  $N_{13}@C_{60}$ & -12678.6 & 39.2\\
    \hline
\end{tabular}
\end{table}
\begin{table}
\textbf{Table II: The Spin multiplicites of $N_{n}@C_{60}$ complex
and difference between spin up and spin down charge densities
}\\

\begin{tabular}
{|c|c|c|c|c|c|c|}\hline
  $N_{n}@C_{60}$ & Q (up-down)& Multiplicity (2S+1)   \\
  \hline
  $N@C_{60} \emph{[N at Center]}$& 3 & 4 [4(expt)]  \\
  $N@C_{60} \emph{[N at the surface bonded]}$ &  1  & 2 \\
  $N_{2}@C_{60}$  & 0 & 1 \\
  $N_{3}@C_{60}$ & 1 & 2 \\
  $N_{4}@C_{60}$ & 0 & 1  \\
  $N_{5}@C_{60}$  & 1 & 2  \\
  $N_{6}@C_{60}$ & 0 & 1   \\
  $N_{7}@C_{60}$  & 1 & 2 \\
  $N_{8}@C_{60}$  & 0 & 1  \\
  $N_{9}@C_{60}$ & 1 & 2  \\
  $N_{10}@C_{60}$  & 0 & 1 \\
  $N_{11}@C_{60}$  & 1 & 2  \\
  $N_{12}@C_{60}$ & 0 & 1  \\
  $N_{13}@C_{60}$ & 3 & 4  \\
    \hline
\end{tabular}
\end{table}
\begin{figure}[figure]
\vspace*{-0.0cm}\centerline{\epsfxsize=6.0in \epsffile{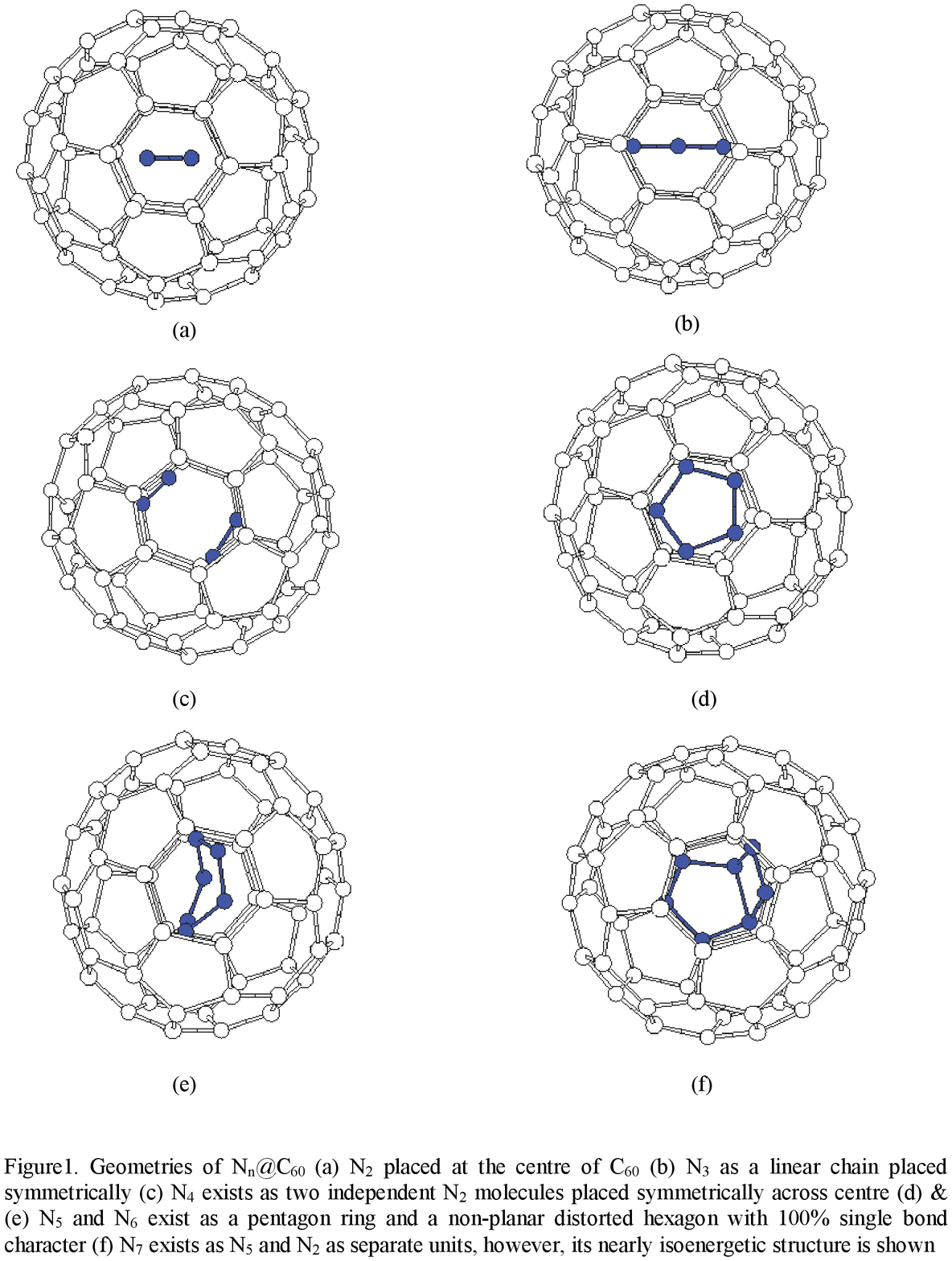}
} \vspace*{-4.0cm} \label{Figure1}
\end{figure}
\begin{figure}[figure]
\vspace*{-0.0cm}\centerline{\epsfxsize=6.0in \epsffile{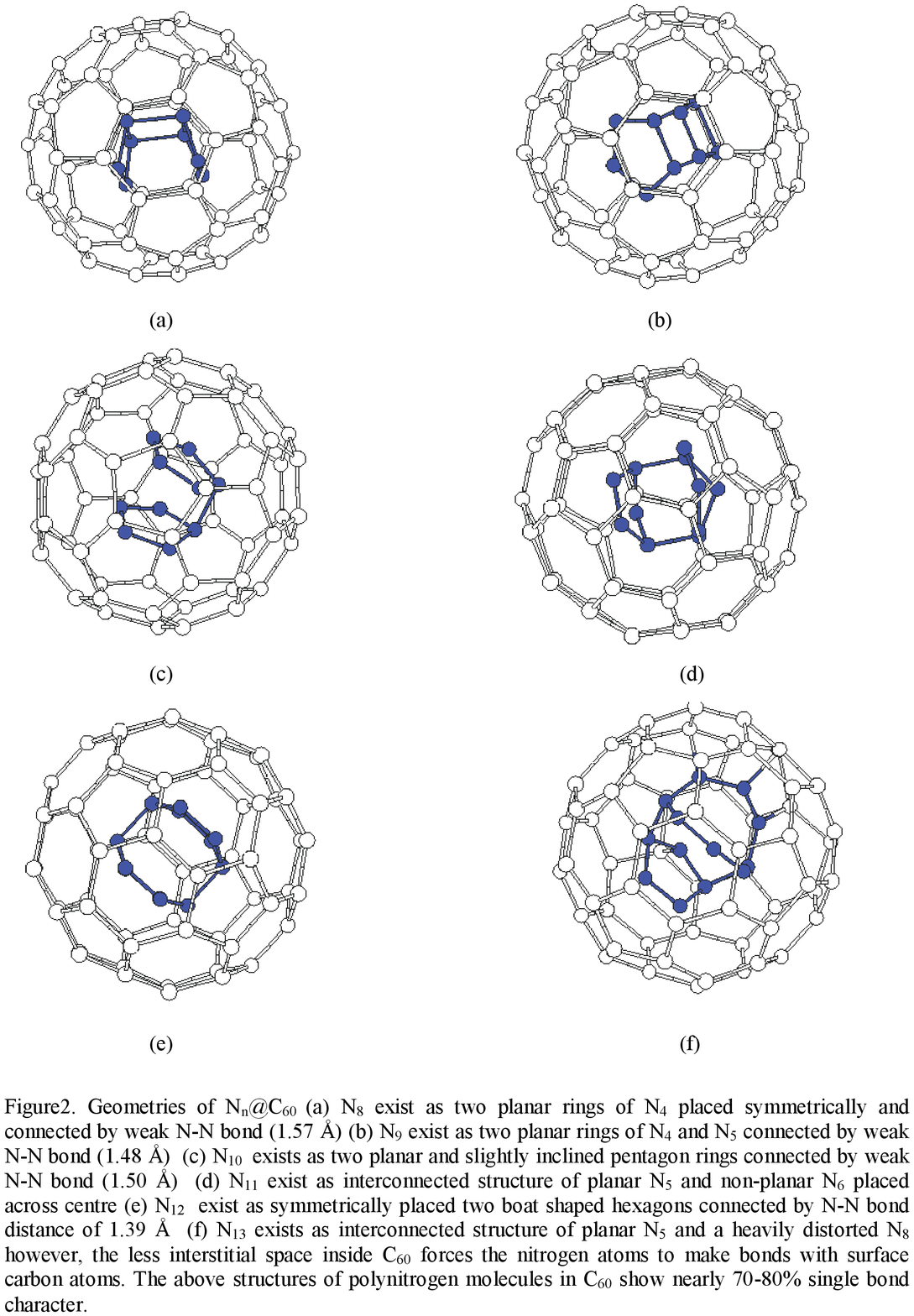}
} \vspace*{-4.0cm} \label{Figure2}
\end{figure}
\begin{figure}[figure]
\vspace*{-0.0cm}\centerline{\epsfxsize=6.0in \epsffile{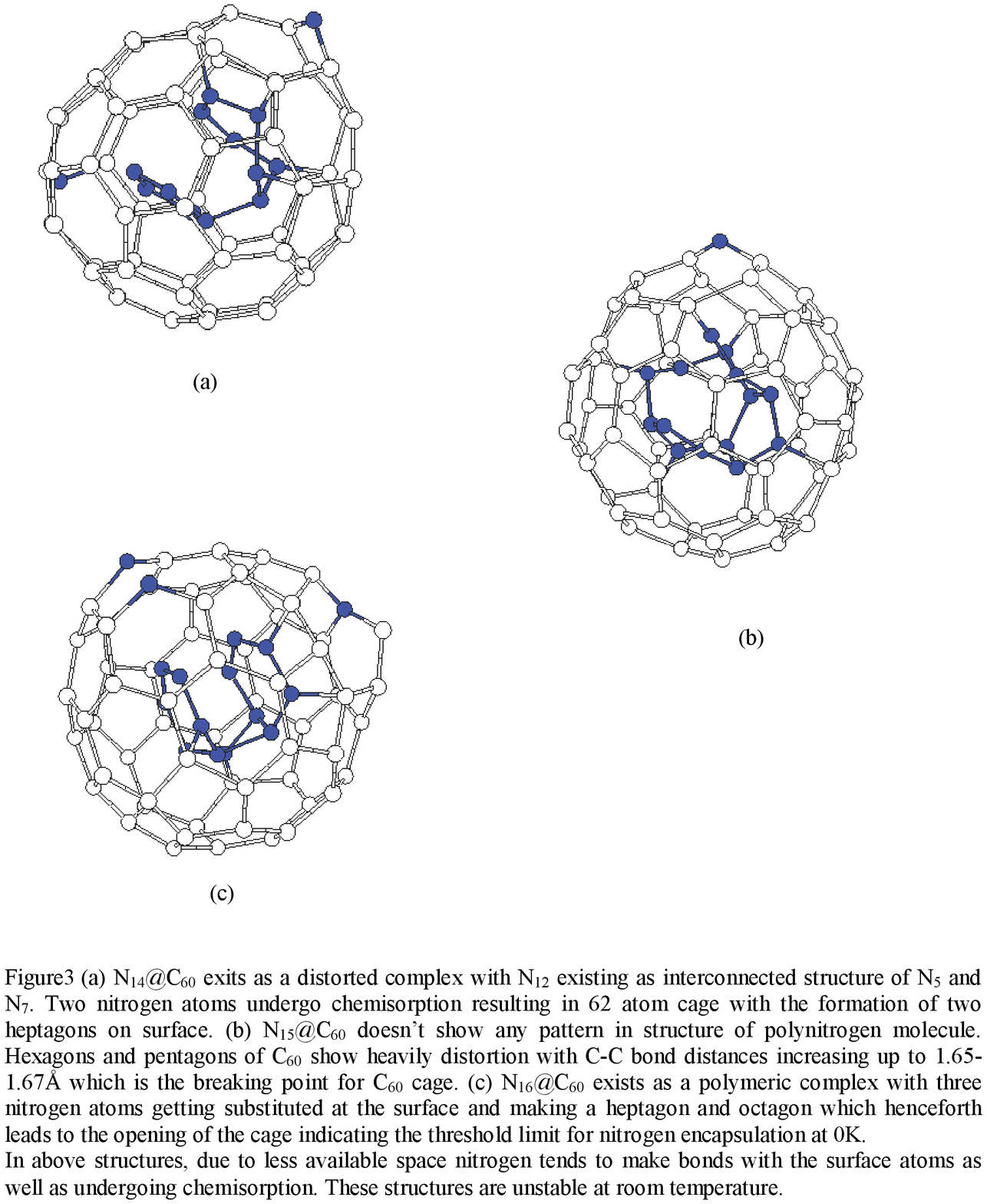}
} \vspace*{-4.0cm} \label{Figure3}
\end{figure}
\begin{figure}[figure]
\vspace*{-0.0cm}\centerline{\epsfxsize=6.0in \epsffile{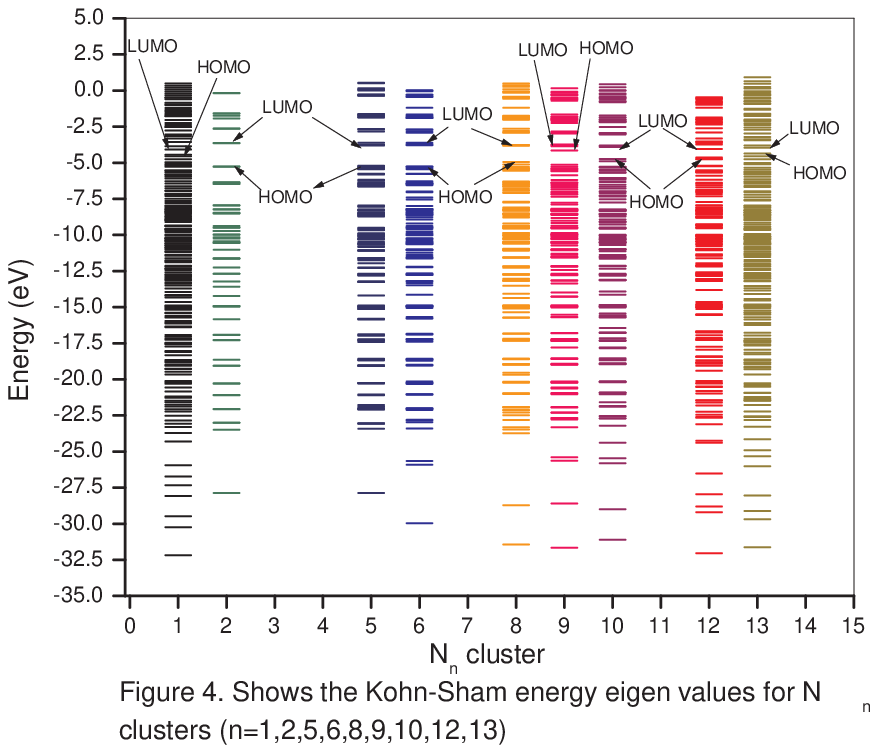}
} \vspace*{-4.0cm} \label{Figure4}
\end{figure}
\begin{figure}[figure]
\vspace*{-0.0cm}\centerline{\epsfxsize=6.0in \epsffile{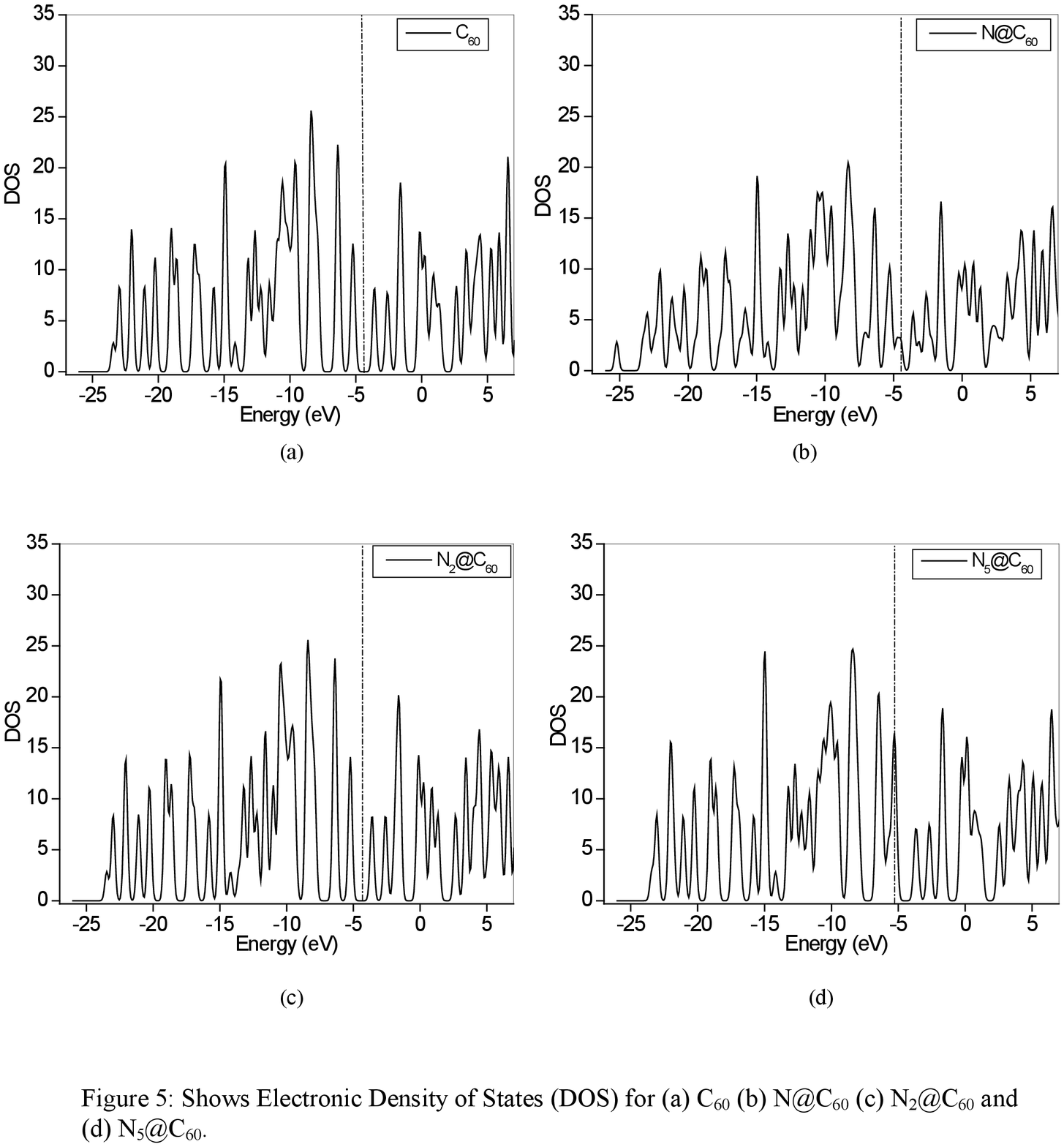}
} \vspace*{-4.0cm} \label{Figure5}
\end{figure}
\begin{figure}[figure]
\vspace*{-0.0cm}\centerline{\epsfxsize=6.0in \epsffile{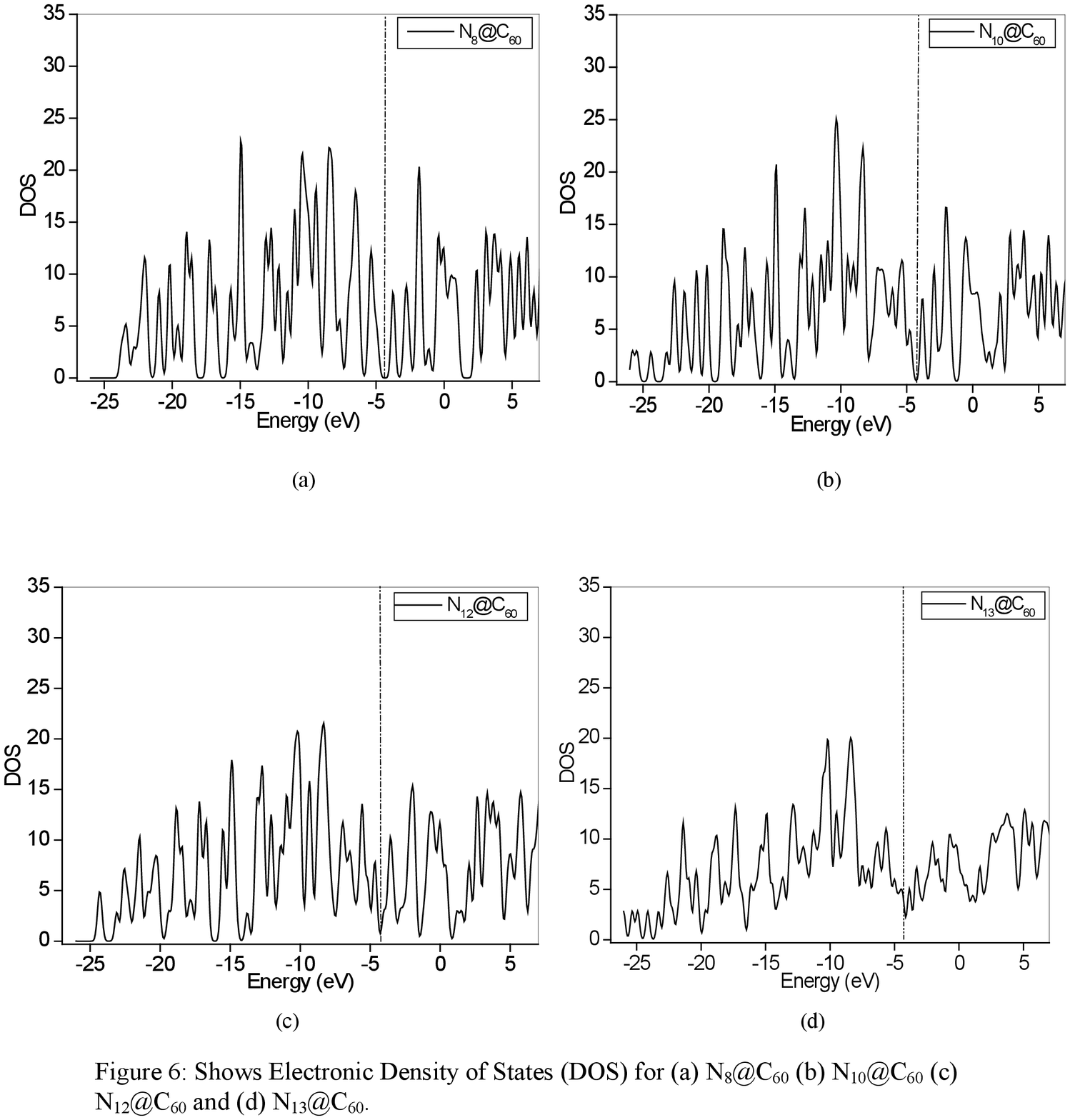}
} \vspace*{-4.0cm} \label{Figure6}
\end{figure}
\begin{figure}[figure]
\vspace*{-0.0cm}\centerline{\epsfxsize=6.0in \epsffile{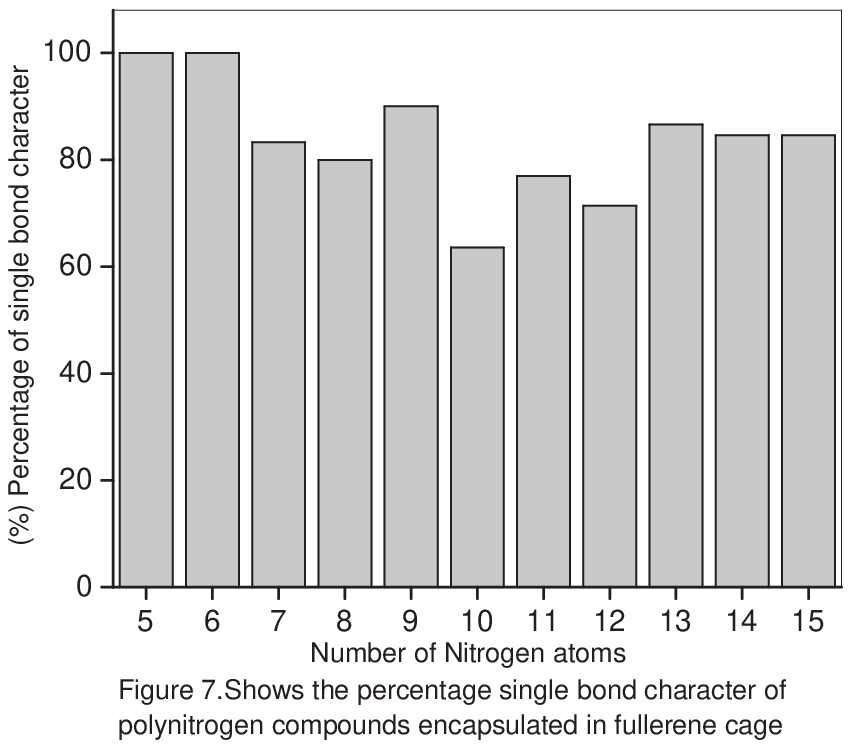}
} \vspace*{-4.0cm} \label{Figure7}
\end{figure}
\begin{figure}[figure]
\vspace*{-0.0cm}\centerline{\epsfxsize=6.0in \epsffile{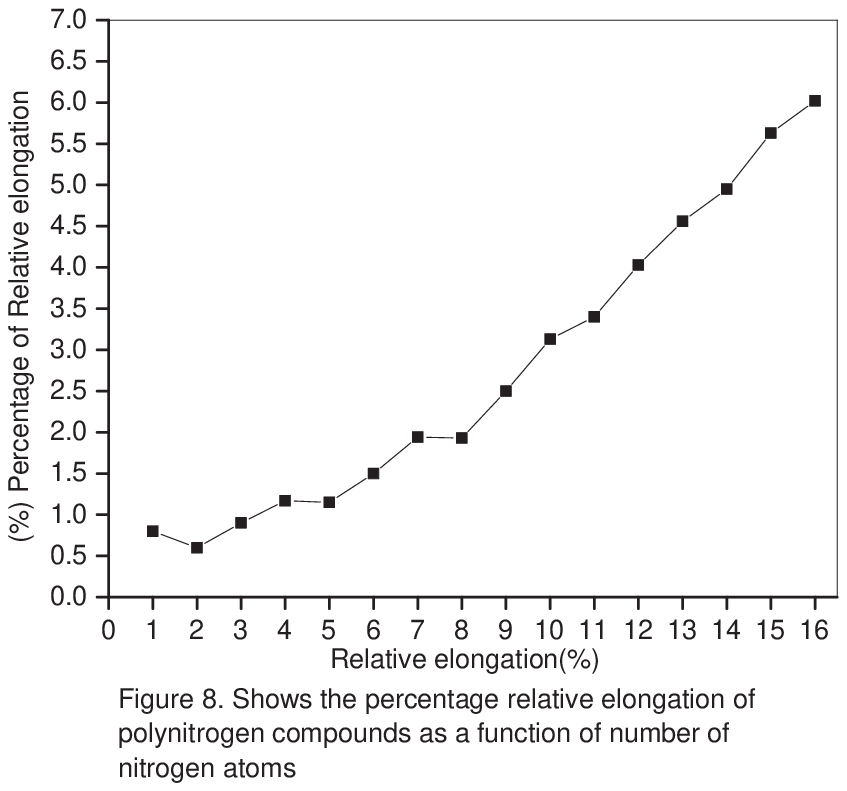}
} \vspace*{-4.0cm} \label{Figure8}
\end{figure}

\begin{thebibliography}{99}
\bibitem{r1} Klapotke TM. Structure and Bonding-High Energy Density Materials. Springer Vol.125, 2007.
\bibitem{r2} Haskins PJ, Fellows J, Cook MD, Wood A. Molecular Level Studies of
Polynitrogen Explosives. Proc.12th Int.Detonation Symp.,California,
2002.
\bibitem{r3} Boates B, Bonev SA. First-Order Liquid-Liquid Phase Transition in
Compressed Nitrogen. Phys Rev Lett 2009;102:015701-4.
\bibitem{r4} Mukherjee GD, Boehler R.  High-Pressure Melting Curve of Nitrogen and the
Liquid-Liquid Phase Transition. Phys Rev Lett 2007;99:225701-4.
\bibitem{r5} Trojan IA, Eremets MI, Medvedev SA, Gavriliuk AG, Prakapenka VB.
Transformation from molecular to polymeric nitrogen at high
pressures and temperatures: In situ x-ray diffraction study Appl
Phys Lett 2008;93:0919071-3.
\bibitem{r6} Zahariev F, Dudiy SV, Hooper J, Zhang F, Woo TK. Systematic Method to New Phases
of Polymeric Nitrogen under High Pressure. Phys Rev Lett 2006;97:
1555031-4.
\bibitem{r8} Christe KO, Wilson WW, Sheehy JA, Boatz JA.
$N_{5}^+$: A Novel Homoleptic Polynitrogen Ion as a High Energy
Density Material. Angew Chem Int Ed 1999;38:2004-9.
\bibitem{r8a} Vij A, Wilson WW, Vij V, Tham FS, Sheehy JA, Christe KO.
Polynitrogen Chemistry. Synthesis, Characterization, and Crystal
Structure of Surprisingly Stable Fluoroantimonate Salts of
$N_{5}^+$. J Am Chem Soc 2001;123:6308-13.
\bibitem{r8b} Shi LW, Chen B, Zhou J-H, Zhang T, Kang Q, Chen
M-B. Structure and Relative Stability of Drum-like $C_{4n}N_{2n}$ (n
= 3-8) Cages and Their Hydrogenated Products $C_{4n}H_{4n}N_{2n}$ (n
= 3-8) Cages. J Phys Chem A 2008;112(46):11724-30.
\bibitem{r8c} Leininger M, Van Huis TJ, Schaefer FH. Protonated High Energy Density Materials:
$N_{4}$ Tetrahedron and $N_{8}$ Octahedron. J Phys Chem Sect A
1997;101:4460-4.
\bibitem{r8d} Ball WD. Tetrazane:
Hartree-Fock, Gaussian-2 and -3, and Complete Basis Set Predictions
of Some Thermochemical Properties of $N_{4}H_{6}$. J Phys Chem Sect
A 2001;105(2):465-70.
\bibitem{r8e} Schulz A, Tornieporth-Oetting IC, Klapotke MT. Nitrosylazid $N_{4}O$,
ein intrinsisch instabiles Stickstoffoxid. Angew Chem Int Ed Engl
1993;32:1610-12.
\bibitem{r8f} Nguyen MT, Ha TK. Azidopentazole is Probably
the Lowest-Energy $N_{8}$ Species. A Theoretical Study. Chem Ber
1996;129:1157-9.
\bibitem{r8g} Fau S, Wilson KJ, Bartlett RJ. On the Stability of
$N_{5}^{+}N_{5}^{-}$. J Phys Chem Sect A 2002;106:4639-44.
\bibitem{r8h} Evangelisti S, Leininger T.  Ionic nitrogen
clusters. J Mol Struct Theochem 2003;621:43-50.
\bibitem{r9} Cacace F, Petris Gde, Troiani A. Experimental Detection of
Tetranitrogen. Science 2002;295:480-1.
\bibitem{r10} Hammerl A, Klapotke TM. Tetrazolylpentazoles: Nitrogen-Rich
Compounds. Inorg Chem 2002;41:906-12.
\bibitem{r10a} Li PC, Guan J, Li S, Quian SL, Wen GX. A theoretical study on the stability of
$N_{15}^+$ cluster. Phys Chem Chem Phys 2003;5:1116–22.
\bibitem{r10b} Cheng L, Li Q Xu W, Zhan S. A computer-aided quantum chemical study of the
$N_{15}^-$ cluster. J Molec Model  2003;9:99-107.
\bibitem{r10c} Liu YD, Yiu PG, Guan J, Li QS. Structures and stability
of $N^{-}11(+)$ and $N^{-}11(-)$ clusters. J Mol Struct (THEOCHEM)
2002;588:37-43.
\bibitem{r10d} Zhou H, Wong N-B, Zhou G, Tian A. What Makes the
Cylinder-Shaped $N_{72}$ Cage Stable? J Phys Chem A 2006;110:7441-6.
\bibitem{r10e} Zhou H, Wong N-B, Zhou G, Tian A. Theoretical
Study on "Multilayer" Nitrogen Cages. J Phys Chem A
2006;110:3845-52.
\bibitem{r10f} Zhou G, Pu X-M, Wong N-B, Tian A, Zhou H. Theoretical Investigation on the Replacement of
CH Groups by N Atoms in Caged Structure (CH). J Phys Chem A
2006;110:4107-14.
\bibitem{r10g} Wang LJ, Zgierski MZ. Super-high energy-rich
nitrogen cluster $N_{60}$. Chem Phys Lett 2003;376:698-03.
\bibitem{r10h} Strout DL. Why Isn't the $N_{20}$ Dodecahedron
Ideal for Three-Coordinate Nitrogen? J Phys Chem A 2005;109:1478-80.
\bibitem{r10i} Wang JL, Lushington GH, Mezey PG. Stability and Electronic
Properties of Nitrogen Nanoneedles and Nanotubes. J Chem Inf Model
2006;46:1965-71.
\bibitem{r10j} Wang L, Mezey PG. Predicted High-Energy Molecules: Helical
All-Nitrogen and Helical Nitrogen-Rich Ring Clusters. J Phys Chem A
2005;109:3241-3.
\bibitem{r10k} Zhao JF, Li QS. Structures and kinetic
stability of $N_{7}$ cluster. Chem Phys Lett 2003;368:12-19.
\bibitem{r10l} Wang LJ, Mezey PG, Zgierski MZ. Stability and the
structures of Nitrogen clusters $N_{10}$. Chem Phys Lett
2004;391:338-43.
\bibitem{r10m} Li QS, Zhao JF. Theoretical Study of Potential
Energy Surfaces for $N_{12}$ Clusters. J Phys Chem A
2002;106:5367-72.
\bibitem{r10n} Wang LJ, Warburton P, Mezey PG. Theoretical
Prediction on the Synthesis Reaction Pathway of $N_{6}$ $(C_{2h})$.
J Phys Chem A 2002;106:2748-52.
\bibitem{r10o} Dixon DA, Feller D, Christe KO, Wilson WW, Vij A, Vij V et al. Enthalpies of Formation of
Gas-Phase $N_{3}$, $N_{3}^-$, $N_{5}^+$, and $N_{5}^-$ from Ab
Initio Molecular Orbital Theory, Stability Predictions for
$N_{5}^{+}N_{3}^{-}$ and $N_{5}^{+}N_{5}^{-}$, and Experimental
Evidence for the Instability of $N_{5}^{+}N_{3}^{-}$.
 J Am Chem Soc 2004;126:834-43.
\bibitem{r10p} Reichlin R, Schiferl D, Martin S, Vanderborgh C, Mills RL. Optical Studies of Nitrogen to 130
GPa. Phys Rev Lett 1985;55:1464-7.
\bibitem{r10q} Bini R, Ulivi L, Kreutz J, Jodl HL. High-pressure
phases of solid nitrogen by Raman and infrared spectroscopy. J Chem
Phys 2000;112:8522-9.
\bibitem{r10r} Olijnyk H,Jephcoat AP. Vibrational Dynamics
of Isotopically Dilute Nitrogen to 104 GPa. \emph{Phys.Rev.Lett.}
1999;83:332-5.
\bibitem{r10s} Mitas L, Martin RM. Quantum Monte Carlo
of nitrogen: Atom, dimer, atomic, and molecular solids. Phys Rev
Lett 1994;72:2438-41.
\bibitem{r10t} Mailhiot C, Yang LH, McMahan AK.
Polymeric nitrogen. Phys Rev B 1992;46:14419-35.
\bibitem{r10u} Lewis SP, Cohen ML. High-pressure atomic phases
of solid nitrogen. Phys Rev B 1992;46:11117-20.
\bibitem{r10v} Martin RM, Needs RJ. Theoretical study
of the molecular-to-nonmolecular transformation of nitrogen at high
pressures. Phys Rev B 1986;34:5082-92.
\bibitem{r11} Eremets MI, Gavriliuk AG, Trojan IA, Dzivenko DA
Boehler R. Single-bonded cubic form of nitrogen. Nature Mater
2004;3:558-63.
\bibitem{r11a} Mattson WD, Sanchez-Portal D, Chiesa S, Martin RM
Prediction of New Phases of Nitrogen at High Pressure from
First-Principles Simulations. Phys Rev Lett  2004;93:125501-4.
\bibitem{r12} Abou-Rachid H, Hu A, Timoshevskii V, Song Y, Lussier
LS. Nanoscale High Energetic Materials: A Polymeric Nitrogen Chain
$N_{8}$ Confined inside a Carbon Nanotube. Phys Rev Lett
2008;100:196401-4.
\bibitem{r12a} Bajwa N, Ingale A, Avasthi DK, Ravi Kumar, Tripathi A, Dharamvir K, Jindal
VK. Role of Electron Energy Loss in Modification of $C_{60}$ Thin
Films by Swift Heavy Ions. J Appl Phys 2008;104:054306-13.
\bibitem{r12b} Kaur N, Gupta S, Dharamvir K, Jindal VK. Formation of dimerized molecule of $C_{60}$ and their solids.
Carbon 2008;46:349-58.
\bibitem{r12c} Kaur N, Dharamvir K, Jindal VK. Dimerisation and fusion of two $C_{60}$ bucky
balls. Chem Phys 2008;344:176-84.
\bibitem{r12d} Kaur N, Gupta S, Dharamvir K, Jindal VK. Behaviour of Bucky ball under extreme internal and external pressure
Proceedings of the 26th International Symposium on Shock Waves
ISSW26,Germany,Springer-Verlag 2007.
\bibitem{r12e} Waiblinger M, Lips K, Harneit W, Weidinger A. Thermal stability
of the endohedral fullerenes $NaC_{60}$ , $NaC_{70}$ , and
$PaC_{60}$. Phys Rev B 2001;63:045421-5.
\bibitem{r12f} Weiden N, Goedde B, KaB H, Dinse K-P.
Squeezing of Nitrogen Atomic Orbitals in a Chemical Trap. Phys Rev
Lett 2000;85:1544-7.
\bibitem {r13} Hebard AF, Rosseinsky MJ, Haddon RC, Murphy DW, Glarum SH, Palstra TTM et al.
Superconductivity at 18 K in potassium-doped $C_{60}$. Nature
1991;350:600-1.
\bibitem{r14} Bethune DS, Johnson RD, Salem JR, de Vries MS, Yannoni CS.
Atoms in carbon cages: the structure and properties of endohedral
fullerenes. Nature  1993;366:123-8.
\bibitem{r15} Sanville E, BelBruno JJ. Computational Studies of Possible Transition
Structures in the Insertion and Windowing Mechanisms for the
Formation of Endohedral Fullerenes. J Phys Chem B 2003;107:8884-9.
\bibitem{r16} Closiowski J, Nanayakkara A. Endohedral fullerites:
A new class of ferroelectric materials. Phys Rev Lett
1992;69:2871-3.
\bibitem{r17} Dietel E, Hirsch A, Pietzak B, Waiblinger M, Lips K,
Weidinger A et al. Atomic Nitrogen Encapsulated in Fullerenes:
Effects of Cage Variations. J Am Chem Soc 1999;121:2432-7.
\bibitem{r17a} Dinse K-P. EPR investigation of atoms in chemical
traps. Phys Chem Chem Phys 2002;4:5442-7.
\bibitem{r18} Murphy TA, Pawlik T, Weidinger A, Hohne M,
Alcala R, Spaeth J-M. Observation of Atomlike Nitrogen in
Nitrogen-Implanted Solid $C_{60}$. Phys Rev Lett 1996;77:1075-8.
\bibitem{r19} Knapp C, Dinse K-P, Pietzak B, Waiblinger M, Weidinger A.
Fourier transform EPR study of $N@C_{60}$ in solution. Chem Phys
Lett  1997;272:433-7.
\bibitem{r19a} Peres T, Cao B, Cui W, Khong A, Jr.Cross JR, Saunders
M et al. Some new diatomic molecule containing endohedral
fullerenes. Int J Mass Spectromery 2001;210/211:241-7.
\bibitem{r19b} Slanina Z, Pulay P, Nagase S. $H_{2}$, Ne, and $N_{2}$ Energies of
Encapsulation into $C_{60}$ Evaluated with the MPWB1K Functional. J
Chem Theory Comput  2006;2:782-5.
\bibitem{r19c} Iwamatsu S.-I, Uozaki T, Kobayashi K, Re S, Nagase S, Murata
S. A Bowl-Shaped Fullerene Encapsulates a Water into the Cage. J Am
Chem Soc 2004;126:2668-9.
\bibitem{r19d} Chai Y, Zhang FQ, Wu LJ. A simple way to
$CN_{x}$/carbon nanotube intramolecular junctions and branches.
Carbon 2006;44:687-91.
\bibitem{r19e} Otero G, Biddau G, Sanchez-Sanchez C, Caillard R, Lopez FM,
Rogero C et al. Fullerenes from aromatic precursors by
surface-catalysed cyclodehydrogenation. Nature 2008;454:865-8.
\bibitem{r20} www.siesta.uae.in
\bibitem{r21} Ordejon P, Artacho E, Soler JM. Self-consistent
order-N density-functional calculations for very large systems. Phys
Rev B (Rapid Comm) 1996;53:R10441-4.
\bibitem{r22} Soler JM, Artacho E, Gale JD, Garcia A, Junquera J, Ordejon P et al.
The SIESTA method for ab initio order-N materials simulation. J
Phys: Condens Matter 2002;14:2745-79.
\bibitem{r23} Junquera J, Paz O, Sanchez-Portal D, Artacho E.
Numerical atomic orbitals for linear-scaling calculations. Phys Rev
B 2001;64:235111-19.
\bibitem{r24} Lee C, Yang W, Parr RG. Development of the
Colle-Salvetti correlation-energy formula into a functional of the
electron density. Phys Rev B 1988;37:785-9.
\bibitem{r25} Kleinman L, Bylander DM. Efficacious Form for Model
Pseudopotentials. Phys Rev Lett 1982;48:1425-8.
\bibitem{r26} Sankey OF, Niklewski DJ. Ab initio multicenter
tight-binding model for molecular-dynamics simulations and other
applications in covalent systems. Phys Rev B 1989;40:3979-95.
\bibitem{r27} Johnson RD, Meijer G, Bethune DS. $C_{60}$ has icosahedral
symmetry. J Am Chem Soc 1990;112:8983-4.
\bibitem{r28} Hertel IV, Steger H deVries J, Weisser B, Menzel C, Kamke B, Kamke
W. Gaint Plasmon Excitation in free $C_{60}$ and $C_{70}$ molecules
studied by Photoionization. Phys Rev Lett 1992;68:784-7.
\bibitem{r29} Lichtenberger DL, Jatcko ME, Nebesny KW, Ray CD, Huffman DR, Lamb LD. Valence and core photoelectron spectroscopy of
$C_{60}$, buckminsterfullerene. Chem Phys Lett 1991;176:203-8.
\bibitem{r30} Utterback NG, Miller GH. Ionization of nitrogen
molecules by nitrogen molecules. Phys Rev  1996;124:1477-81.
\bibitem{r31} Mauser H, Nicolaas JR, Hommes van E, Clark T, Hirsch A, Pietzak B, Weidinger
A, Dunsch L. Stabilization of Atomic Nitrogen Inside $C_{60}$. Angew
Chem Int Ed 1997;36:2835-8.
\bibitem{r32} Minh Tho Nguyen. Polynitrogen compounds
1. Structure and stability of $N_{4}$ and $N_{5}$ systems.
Coordination Chem Rev 2003;244:93-113.
\bibitem{r33} Tobita M, Bartlett RJ. Structure and Stability of $N_{6}$ Isomers
and Their Spectroscopic Characteristics. J Phys Chem
2001;105:4107-13.
\bibitem{r34} Sharma H, Garg I, Dharamvir K, Jindal VK. Structural,
electronic and vibrational properties of $C_{60-n}N_{n}$ (n=1-12). J
Phys Chem A 2009;113:9002-13.
\bibitem{r35} Kobayashi K, Nagase S, Dinse K-P. A theoretical
study of spin density distributions ans isotropic hyperfine
couplings of N and P atoms in $N@C_{60}$, $P@C_{60}$, $N@C_{70}$,
$N@C_{60}(CH_{2})_{6}$ and $N@C_{60}(SiH_{2})_{6}$. Chem Phys Lett
2003;377:93-8.
\bibitem{r36} Lu J, Zhang X, Zhao X. Electronic structures of
endohedral $N@C_{60}$, $O@C_{60}$ and $F@C_{60}$. Chem Phys Lett
1999;312:85-90.
\bibitem{r37} Greer JC. The
atomic nature of endohedrally encapsulated nitrogen $N@C_{60}$
studied by density functional and Hartree-Fock methods. Chem Phys
Lett 2000;326:567-72.
\end{thebibliography}
\end{document}